\documentstyle[12pt]{article}
\begin{document}
\setlength{\baselineskip}{24pt}

\newcommand{\bea}{\begin{eqnarray}}
\newcommand{\eea}{\end{eqnarray}}
\newcommand{\be}{\begin{equation}}
\newcommand{\ee}{\end{equation}}

\newcommand{\nue}{\mbox{\(\nu_{\rm e}\)}}
\newcommand{\numu}{\mbox{\(\nu_\mu\)}}
\newcommand{\nutau}{\mbox{\(\nu_\tau\)}}
\newcommand{\munu}{\mbox{\(\mu_\nu\)}}
\newcommand{\mnu}{\mbox{\(m_\nu\)}}
\newcommand{\mnutau}{\mbox{\(m_{\nu_\tau}\)}}
\newcommand{\mnua}{\mbox{\(m_{\nu_a}\)}}
\newcommand{\mea}{\mbox{\(m_{{\rm e}_a}\)}}

\newcommand{\Fvc}{F_{\rm vc}}
\newcommand{\Fdg}{F_{\rm dg}}

\newcommand{\GF}{\mbox{\(G_{\rm F}\)}}
\newcommand{\muB}{\mbox{\(\mu_{\rm B}\)}}
\newcommand{\mPl}{\mbox{\(m_{\rm Pl}\)}}

\newcommand{\mW}{\mbox{\(m_{\rm W}\)}}
\newcommand{\me}{\mbox{\(m_{\rm e}\)}}
\newcommand{\Wangle}{\mbox{\(\theta_{\rm W}\)}}
\newcommand{\pF}{\mbox{\(p_{\rm F}\)}}
\newcommand{\EF}{\mbox{\(E_{\rm F}\)}}
\newcommand{\dgket}{\mbox{\( | {\rm dg}\rangle\)}}
\newcommand{\dgbra}{\mbox{\( \langle {\rm dg}| \)}}

\newcommand{\EW}{\mbox{\(E_{\rm W}\)}}
\newcommand{\Ee}{\mbox{\(E_{\rm e}\)}}
\newcommand{\Enu}{\mbox{\(E_nu\)}}

\newcommand{\sproduct}[2]{\left( #1 #2 \right)}
\newcommand{\sigmabar}{\bar \sigma}
\newcommand{\xbar}{\bar x{}}
\newcommand{\Abar}{\bar A{}}
\newcommand{\vevof}[1]{\langle 0| #1 |0\rangle}
\newcommand{\T}{\mbox{T}}
\newcommand{\TR}{\mbox{T}}
\newcommand{\PT}{\mbox{P}}
\newcommand{\CC}{\mbox{C}}

\newcommand{\ec}{\mbox{\( \rm e^{\rm c}\)}}
\newcommand{\e}{\mbox{e}}

\newcommand{\hc}{\mbox{(h.c.)}}

\title{
Enhancement of the Transition Magnetic  Moments of a Neutrino
by   Degenerate  Electrons }
\author{\vspace*{0.5em}\\
Hisashi Kikuchi \\
\vspace*{0.5em}\\
{\normalsize \sl Ohu University}\\
{\normalsize \sl Koriyama, 963 Japan}
}
\date{OHU-PH-9510}

\maketitle

\vfil

\begin{abstract}
The one-loop induced magnetic dipole moments of a neutrino are examined in
a background of degenerate electrons in the standard model.
For the  nonrelativistic neutrino, they are enhanced by a factor
\( (8\pF/3\mnu) \), where \pF\ is the electron Fermi momentum and
\mnu\ the neutrino mass.
For the relativistic neutrino, they  enhance  the flavor-changing
but helicity-conserving  process  because of the absence of
the GIM cancellation.
\end{abstract}
\vfil

\newpage

\section{Introduction}
The electromagnetic interaction of a neutrino in a hot and/or dense medium
has been an interesting subject   both in particle physics and astrophysics.
Adams, Ruderman, and Woo
calculated the decay rate of a plasmon into a pair of (anti-)neutrinos
in a degenerate electron plasma that is realized in stellar interiors
\cite{ada}.
The plasmon decay is by now shown to be the dominant cooling process
in  dense stars \cite{ada,bea,sut}.
If a neutrino has  magnetic dipole moments,   of diagonal and/or
(flavor-changing) transition
type, it undergoes a spin  rotation when travelling in magnetic field
for both cases of Dirac \cite{fuj} and Majorana particle \cite{sch}.
This spin rotation was  studied intensively in the context that it provides
a  solution to the solar neutrino problem  \cite{vol,lim}.
These  ideas have motivated the study on the  electromagnetic
vertex of a neutrino in a medium in the standard model
\cite{ora,nie,dol,giu,alt,mas} a
Especially Nieves, D'Olivo and Pal have shown that the result leads us to
the drastic enhancement of the radiative decay rate of a neutrino \cite{dol},
which is strongly suppressed by the GIM cancellation in the vacuum \cite{lee}.
Subsequently, Giunti, Kim, and Lam have  shown this enhancement can also
be understood as  the coherent scattering of a neutrino off
the background electrons \cite{giu}.
They further argued that there is no significant enhancement of
the neutrino  magnetic  moments   \cite{giu}.

In this paper, we examine  the magnetic dipole  moment induced by the
one-loop processes that have been studied by Nieves et al.
(depicted in Fig.~1 and 2)  especially for the case of degenerate electrons,
and show it gets  in fact { \em enhanced}.
The  magnetic moment we  find is of the magnitude of
\(\munu \sim e \GF n_{\rm e}^{1/3}\) where e is the electromagnetic
coupling constant, \GF\  the Fermi coupling constant,
and \( n_{\rm e}\) the electron density.   Thus it is consistent
with the assertion of Giunti et al.  that there is no enhancement proportional
to \(e \GF n_{\rm e}\),  but it  is still much larger than their
speculation that the result would be  even smaller than  the vacuum
contribution.%
\footnote{The one-loop effect  we calculate   includes the same
contribution calculated by Giunti et al.   in Ref.~\cite{giu}
although the diagrams appear
different:  according to the standard formalism for many particle systems,
we can forget about the existence of background electrons
if we use  the appropriate propagators for  internal electron lines
 in the perturbative expansion \cite{fet};
since the external electron lines in Fig.~2 in  their paper represent
the background electrons,  those diagrams correspond to  Fig.~1~(a) in
our language.}

The induced magnetic moments are  helicity-conserving for  relativistic
neutrinos.
This is due to the left-handed nature of the interaction in the standard
model.
Thus they will hardly  generate  the spin-rotation for relativistic
neutrinos,
which is thought to be a main consequence of nonzero magnetic moments.
They can, however,   enhance  flavor-changing processes under the
presence of magnetic field.
This is because of the absence of the GIM cancellation in the degenerate
electron plasma: the one-loop induced
magnetic moments  in the vacuum are independent of the internal
charged lepton mass to the leading order in the Fermi coupling constant
and, thus, cancel out  when one adds all the contribution from
the three species \cite{lee};
in the electron background, only electrons give extra contribution
proper to their degeneracy and thus the GIM cancellation no more works.

We notice that the magnetic moment we evaluate in this paper is
the same one as  evaluated by Paster, Semikoz, and Valle \cite{pas}
in the study of transitions between the standard neutrinos and
a sterile one,
although their context and method of calculation are different
from ours.
In their paper, they claimed that the moments can only affect
the dispersion relation of neutrinos.
We would like to emphasize that they can cause a flavor-changing process
as well if the mass difference of the neutrinos are big enough and  mass
eigenstates do not loose their identity even under
the presence of coherent scattering in degenerate electrons.

We simply adopt the zero temperature approximation.
The result is thus applicable to a plasma whose Fermi momentum
\pF\ is much larger than its temperature \(T\).
This condition reads
\be [  Y_{\rm e}\,\rho( {\rm g/cm^3})] ^{1/3} \gg
1.7 \times 10^{-8} \, T(\rm K)\ee
if one uses
\(\pF = (3\pi^2 n_{\rm e})^{1/3} \) and
 \(n_{\rm e}({\rm 1/cm^3}) \simeq 6.0\times 10^{23} \, Y_{\rm e}
 \rho( {\rm g/cm^3})\),
where \(n_{\rm e}\) is  the electron number density (in units of 1/cm\(^3\)),
 \(\rho\) the mass density (in units of  g/cm\(^3\)), and \(Y_{\rm e}\)
the electron fraction per baryon.
This condition is satisfied in the stellar interiors  such as
the center of the Sun and the core of a supernova at the onset of collapse
\cite{suz}.

We do the calculations in the left-handed two component notation;
the general structure of the neutrino electromagnetic vertex is more
clearly represented in this notation than in
the regular four component Dirac notation.
In section 2 we  provide the basic formula for the calculation
and specify  the electromagnetic vertex in the vacuum  by applying
discrete symmetries, C, P, and T.
In section 3, we will calculate the magnetic dipole moment in a degenerate
electron plasma and discuss its physical consequence.

\section{Electromagnetic vertex in the two-component notation}

The Lorentz group has two different spinorial (two-dimensional)
representations,  called left-handed and right-handed.
Each  is  the complex conjugate of the other.
We assume  spinor field operators,  e.x.
\(l(x)\),  obey the left-handed representation,
and thus their hermitian conjugates,
 \( \epsilon \l^\dagger(x)\),\footnote{%
Notations:  `Dagger' on the field operators means the hermitian conjugate
in the Hilbert space; it  does not include the operation of transposition
which is usually assumed for the column of two-component field operators.
Products of two field operators such as \( (\epsilon l_a l_a) \) is understood
as
\( (\epsilon)_{\alpha\beta} (l_a)_{\beta} (l_a)_\alpha \) if one explicitly
writes  the indices \(\alpha , \beta , ... = 1, 2\)  in the spinor space.}
obey the right-handed one,
where \(\epsilon \equiv i\sigma^2, \) is defined with
the second component of the Pauli matrices \(\vec \sigma\).
The most general form of the free Lagrangian  for \(n\) species of fields,
\(l_a(x)\)
(\(a = 1, ..., n\)), is written as
\be
{\cal L}_{\rm M} = \sum_a \left[ l_a^\dagger i( \sigmabar^\mu \partial_\mu)
 l_a -
{1\over2}
m_a \sproduct{ \epsilon l_a } {l_a} - {1\over 2} m_a \sproduct{ l_a^\dagger}
{ \epsilon l_a^\dagger } \right],
\label{LforM}\ee
where  \( \sigmabar^\mu \equiv ( 1, -\vec\sigma\, ) \) and
\(m_ a\) are the real positive mass parameters.
One may   write the mass term in a matrix form
 \( M_{ab}  l_a \epsilon l _b /2  + \hc \)
with a symmetric matrix \(M_{ab}\)
(note \( l_a \epsilon l_b =  l_b \epsilon l_a \)).
All the parameters in \(M_{ab}\), however, are not physically relevant since
it can always be diagonalized by the field redefinition,
\(l_a \rightarrow U^{-1}_{ab}l_b\),  with
the unitary matrix  \( U_{ab}\) that satisfies
\be M_{ab} = \sum_c m_c U_{ca} U_{cb}. \ee
If there is no degeneracy in  \(m_a\),
the Lagrangian (\ref{LforM}) describe  Majorana spinors.

We assume all the neutrinos \nue, \numu, and \nutau\
are Majorana fermions   and their free part of
the Lagrangian has the same form as (\ref{LforM}).
We use the subscripts e, \(\mu\), and \(\tau\) for distinguishing the
mass eigenstates.
After the  quantization, they  are expanded
in  the annihilation  operator \(a_{\nu_a}(\vec p, s)\) and
the creation operator
\(a^\dagger_{\nu_a}({\vec p, s})\) of the one-particle state with
momentum \(\vec p\) and helicity \(s, (s=\pm1)\).
They are collectively written as
\be
\nu(x) = \int {d \vec p\over (2\pi)^{3\over 2} } \sum_s
\left[ e^{ -i E_\nu(\vec p\,) x^0 + i \vec p\cdot \vec x } u(\vec p, s) \,
a_\nu({\vec p, s})
+ e^{iE_\nu(\vec p\,) -i \vec  p\cdot \vec x} v(\vec p, s) \,
a^\dagger_\nu({\vec
p, s}) \right] \label{nuex}
\ee
where \(E_\nu(\vec p\,) \equiv \sqrt{ \vec p\,^2 + m_\nu ^2 }\)
is the energy of the neutrino.
The spinor functions  \(u(\vec p, s)\) and \(v(\vec p, s)\) in Eq.~(\ref{nuex})
are given by
\be u(\vec p, s) = \sqrt{ E_\nu - s |\vec p\,| \over 2E_\nu } \chi( \hat p, s
),
\quad
\epsilon\, v^*( \vec p, s ) = \sqrt{ E_\nu + s |\vec p\,| \over 2E_\nu }
\chi( \hat p, s ), \label{uv}\ee
where \(\chi(\hat p, s) \) is  the  helicity eigenspinor defined by
\be (\vec \sigma \cdot \hat p\, ) \; \chi(\hat p, s) = s \chi(\hat p, s ),
\quad
\chi(\hat p, s) ^\dagger \chi(\hat p, t)  = \delta_{st}. \ee
They have been obtained by
requiring that they must obey a relation
\be (p_\mu\sigmabar^\mu) u(\vec p, s) = m_\nu \epsilon v^*(\vec p, s) \quad
(p^0 = E_\nu(\vec p\,) )\ee
that comes from the equation of motion
\(i\partial_\mu \sigmabar^\mu\, \nu - m_\nu \epsilon \, \nu^\dagger = 0
\label {Maj}\)
and that the canonical commutation relations of the spinor-components
\(\nu_\alpha(x)\) (\(\alpha = 1, 2\)) of the field operators,
\be \{ \nu_\alpha (\vec x\,), \nu_\beta^\dagger(\vec y\,) \}
= \delta(\vec x -\vec y\,) \delta_{\alpha \beta}, \quad
\{ \nu_\alpha (\vec x\,), \nu_\beta (\vec y\,) \} = 0, \quad
\{ \nu_\alpha^\dagger (\vec x\,), \nu_\beta^\dagger (\vec y\,) \} = 0, \ee
result in the properly normalized commutation relations
\be \{ a^\dagger _\nu({\vec p,s}), a_\nu({\vec q, t}) \} = \delta(\vec p - \vec
q\,)
\delta_{st}, \quad
\{ a_\nu({\vec p,s}), a_\nu({\vec q, t}) \} = 0, \quad
\{ a_\nu^\dagger({\vec p,s}), a_\nu^\dagger({\vec q, t}) \} = 0. \ee

The charge conjugation  C and parity transformation  P are unitary operators
in the Hilbert space,  and  the time reversal  T is an anti-unitary operator
\cite{bjo}.  They transforms the neutrino fields as
\bea
\mbox{C}: \nu(x)  &\longrightarrow&  \nu(x), \label{CC}\\
\mbox{P}: \nu(x ) &\longrightarrow & i \epsilon \nu^\dagger(\bar x),
\label{PT}\\
\mbox{T}: \nu(x ) &\longrightarrow & \epsilon \nu (-\bar x)
\label{TR}, \eea
where \(\bar x^\mu = (x^0, - \vec x\,)\).
We will use the notation that `bar' on a four vector means reversing the
direction of its spatial components while keeping the temporal component
unchanged.
The free Lagrangian for \(\nu(x)\)  is symmetric under all these
discrete symmetries.
Note  that T replaces  all the c-numbers  with their complex
conjugates as well \cite{bjo}.

The charged leptons  e, \(\mu\), and \(\tau\) are Dirac particles and
described by a pair of two-component fields  with a degenerate mass.
Assume \(l_1\) and \(l_2\) have the same mass in (\ref{LforM}).
Then the  Lagrangian  possesses an O(2) symmetry.
We redefine the fields as
\be \e \equiv {1\over \sqrt 2} (l_1 + i l_2)   \quad
   \ec \equiv  {1\over \sqrt 2} (l_1 - i l_2)
\ee
and  identify the O(2) to the electromagnetic  U(1)  gauge symmetry.
Switching on the coupling  to the electromagnetic field \(A_\mu\),
we write the  Lagrangian as
\be
{\cal L}_{\rm e} = \e^\dagger i \sigmabar^\mu(\partial_\mu - ie A_\mu) \e
+\ec^\dagger i \sigmabar^\mu (\partial_\mu + ie A_\mu ) \ec
- \me  \sproduct {\epsilon \e}{\ec}
- \me \sproduct{\e^\dagger}{\epsilon\ec^\dagger}.
\label{QED}\ee
After quantizing its free part,
we get
\bea
\e(x) = \int {d \vec p\over (2\pi)^{3 \over2} } \sum_s
\left[ e^{ -i E_{\rm e}(\vec p\,) x^0 + i \vec p \cdot \vec x } u(\vec p, s)\,
a_{\rm e}({\vec p, s})
+ e^{ i E_{\rm e}(\vec p\,) x^0 -i \vec p\cdot \vec x} v(\vec p, s) \,
b_{\rm e}^\dagger({\vec p, s}) \right], \\
\ec(x) = \int {d \vec p\over (2\pi)^{3\over 2} } \sum_s
\left[ e^{ -i E_{\rm e}(\vec p\,) x^0 + i \vec p \cdot \vec x} u(\vec p, s) \,
b_{\rm e}({\vec p, s})
+ e^{i E_{\rm e}(\vec p\,) x^0 -i \vec p\cdot \vec x } v(\vec p, s) \,
 a^\dagger_{\rm e}({\vec p, s}) \right],
\eea
where  operators \(a_{\rm e}(\vec p,s) \) stand for electrons and
\(b_{\rm e}(\vec p,s) \) for  positrons;
 \(u(\vec p,s)\) and \(v(\vec p,s)\) are defined by the same expressions as
(\ref{uv}) using the electron energy
 \(E_{\rm e}(\vec p\,) \equiv \sqrt{ \vec p\,^2 + m_{\rm e}^2 }\)
instead of \(E_\nu\).

The discrete symmetries transform the fields as\footnote{
 The definitions of C, P, and  T have the ambiguities of a sign for the
Majorana neutrinos and of an U(1) phase for the charged leptons.
These ambiguities   are eventually  fixed by  the interaction terms.
The definition of P here is different from the usual one \cite{bjo};
the intrinsic parity of all fermion is (\(-i\))  in our definition.}
\be \CC:\; \e(x) \longleftrightarrow \ec(x);
\; A_\mu(x) \longrightarrow - A_\mu(x), \label{eCC}\ee
\be\PT: \left\{ \begin{array}{ccc}
\e(x) &  \longrightarrow & i \epsilon \ec^\dagger(\bar x) \\
\ec(x) &  \longrightarrow & i \epsilon \e^\dagger(\bar x)
\end{array}\right. ;A_\mu(x) \longrightarrow \bar A_\mu(\bar x),\label{ePT}\ee
and
\be \TR:\left\{ \begin{array}{ccc}
\e(x) &  \longrightarrow &  \epsilon \e(-\bar x) \\
\ec(x) &  \longrightarrow &  \epsilon \ec(-\bar x)
\end{array}\right. ;A_\mu(x) \longrightarrow \bar A_\mu(-\bar x).
\label{eTR}\ee

Using these explicit expansions of the field operators, we calculate
the one-loop induced electromagnetic vertex  of a neutrino
with the electromagnetic interaction of the charged leptons and
the interactions of  the massive vector bosons \(W^{\pm}\),
of the form
\bea
{\cal L}_{\rm int}& = &
- {g\over\sqrt 2}\left[ V_{\nu_ab} \nu_a^\dagger  \sigmabar^\mu \e_b
W^+_\mu
+ V^*_{\nu_b a} \e^\dagger_a \sigmabar^\mu \nu_b W^-_\mu \right]
\nonumber \\
&&- e \left[  ( i F_{\mu\nu}) W^+{}^\mu W^-{}^\nu + (i W^+_{\mu\nu}) W^-
{}^\mu A^\nu + (i W^-_{\mu\nu}) A^\mu W^+{}^\nu \right],
\eea
where  the indices \(a, b \) ( = \e, \(\mu, \tau\) ) are used to distinguish
the
flavors,
\(V_{\nu_a b} \) is the element of the CKM  matrix between
\(\nu_a\)  and the charged lepton \(\e_b\).
The relevant Feynman diagrams are depicted in Fig.~1.
We will  not explicitly calculate the contribution from
the  neutral current interaction.
It   does not contribute to transition  moments  in the vacuum or in an
electron background.
It contributes to  diagonal magnetic moments in the electron
background through the process depicted in Fig.~2.
The correction from it is easily made,
as we will show  in Section 3.

The W-bosons  are defined to transform as
\be
\mbox{CP}: W_\mu^\pm(x) \rightarrow - \bar W_\mu^\mp(\bar x)
\label{wCP}\ee
and
\be
\TR: W_\mu^\pm(x) \rightarrow \bar W_\mu^\pm(-\bar x)
\label{wTR}\ee
under CP and T.
The whole Lagrangian, the sum of \({\cal L}_{\rm int}\),
\({\cal L}_{\rm M}\) for the neutrinos,
 \({\cal L}_{\rm e}\) for the charged leptons, and  the kinetic terms for
the vector bosons, is CP- and
T-invariant up to complex phases in \(V_{\nu_a b}\).

Let us write the effective interaction induced by  electron  intermediate
state as
\be {\cal L}_{\rm eff} = \left( {- e g^2\over 2 }\right)
V_{\nu_a \rm e} V^*_{\nu_b \rm e} \nu_a^\dagger (p_1) F^\mu(p_1,p_2)
\nu_b(p_2)
A_\mu(q), \label{Leff}\ee
where the field operators  are written in their Fourier components
of momenta \(q\), \(p_1\), and \(p_2\); our convention is that their directions
 are inward to the vertex and thus  \(q + p_1+ p_2 = 0\).
The vertex function \(F^\mu\)  is evaluated by
\bea F^\mu(p_1, p_2) &\equiv& \int\int dx dy e^{ -i p_1 x -i p_2 y }
\nonumber\\
&&\times
\langle\:| \, {\rm T}\,  W^+_\nu(x) \bar\sigma^\nu \e(x) \, \left[ e^\dagger(0)
\bar\sigma^\mu e(0)
- \ec^\dagger(0) \bar\sigma^\mu \ec(0) + ... \right] \,
\e^\dagger(y) \bar\sigma^\lambda W^-_\lambda (y) |\:\rangle,\label{Fmu}
\nonumber\\
\eea
where T here denotes the time-ordered product and \(|\:\rangle\) is the
vacuum
or the groundstate of a electron gas depending on which background we
are considering. We have abbreviated the W's electromagnetic vertex
in (\ref{Fmu}).

We first consider \(\Fvc^\mu\), the vertex function in the vacuum, and
see how the magnetic dipole moments look like in the two-component
notation.
There are six independent structures for \(\Fvc^\mu\);
\( \sigmabar^\mu\), \((\sigmabar p_1) p_1^\mu\),
\((\sigmabar p_2) p_2^\mu\),  \((\sigmabar p_1) p_2^\mu\),
 \((\sigmabar p_2) p_1^\mu\),  and
\(\epsilon^{\mu\nu\lambda\kappa} \sigmabar_\nu p_{1\lambda} p_{2\kappa}\),
where \(\epsilon^{\mu\nu\lambda\kappa}\) is the four dimensional
anti-symmetric tensor (\(\epsilon_{0123} = - \epsilon^{0123} = 1\)).
The  vacuum is CP- and T-invariant.  This   leads us to the identities
that \(\Fvc^\mu\) must obey;
inserting (CP)\(^{-1}\)CP and T\(^{-1}\)T into Eq.~(\ref{Fmu})
and transforming the fields operators according to (\ref{eCC})--(\ref{eTR}),
(\ref{wCP}), and (\ref{wTR}), we obtain
\be \Fvc^\mu(p_1, p_2) = - \epsilon \bar \Fvc^\mu{}^{\rm T}(\bar p_2, \bar
p_1)
\epsilon \quad \mbox{(by CP)}\ee
 and
\be \Fvc^\mu(p_1, p_2) = - \epsilon \bar \Fvc^\mu{}^{\rm T}(- \bar p_2, -
\bar p_1)
\epsilon \quad \mbox{(by T)}, \label{Tinv}\ee
where `T' on a matrix denotes the transposition.
\(\Fvc^\mu\) also satisfies the gauge invariance
\be q_\mu \Fvc^\mu = 0. \ee
These constraints reduce the number of form factors to two,
\be
\Fvc^\mu (p_1, p_2) = f_{\rm T}\, \left( q^\mu q^\nu - q^2 g^{\mu\nu}\right)
\sigmabar_\nu
+ i f_{\rm M}\, \epsilon^{\mu\nu\lambda\kappa} \sigmabar_\nu
 p_{1\lambda} p_{2\kappa}. \label{Fvc}
\ee
The form factors \(f_{\rm T}\) and  \(f_{\rm M}\)
are functions of the Lorentz invariants \(p_1^2\),  \(p_2^2\),
and \(( p_1 p_2) \), as well as the mass parameters \me\ and \mW.
The second term in (\ref{Fvc}) contains   magnetic dipole moments.
For nonrelativistic momenta, \(|\vec p_1\,|, |\vec p_2\,| \ll m_{\nu_a},
m_{\nu_b}\),
it is dominated by the spatial components and contains the element
\be
 i f_{\rm M}\, {m_{\nu_a}+m_{\nu_b} \over 2}
\left[\vec\sigma\times (- \vec p_1 - \vec p_2\,) \right]. \ee
This is readily translated to the effective coupling to the magnetic
field \(\vec B = \vec\nabla\times \vec A\)
of the form
\be {\cal L}_{\rm m} = \mu_{ab} (\nu_a^\dagger\vec \sigma \nu_b ) \cdot
\vec B, \label{Lm}\ee
where
\be
\mu_{ab} \equiv { e g^2 \over 2} V_{\nu_a {\rm e}} V^*_{\nu_b {\rm e}}
f_{\rm M} {m_{\nu_a}+ m_{\nu_b} \over 2}. \ee
To leading order in \( \me^2/\mW^2 \), \(f_{\rm M}\) is independent of \me\
and
\be
f_{\rm M}  = {3 \over 32 \pi^2 \mW^2}, \label{fM}\ee
which is obtained from  the result in Ref.~\cite{lee}.
Since intermediate muon  and tauon  give the same contribution
and \(V_{\nu_a b}\) is an unitary matrix,
the magnetic dipole moments induced in the vacuum are
of diagonal type of the magnitude
\be \munu \equiv \mu_{aa} (\mbox{no summation over \it a}) =
{3 e \GF \mnu \over 8\sqrt 2 \pi^2} \label{muvc}
\ee
and the transition moments, \(a\neq b\), vanish to the leading order
(GIM cancellation) \cite{lee}.

Note also that the matrix element of \({\cal L}_{\rm m}\) between
nonrelativistic
Majorana neutrinos  is  further suppressed than the case where they
were of Dirac type.
This is readily verified by  inserting the expansion
 (\ref{nuex})  into (\ref{Lm}).
[The expansion for the Dirac neutrino is given by replacing
\(a_\nu^\dagger\) with \(b_\nu^\dagger\), the creation operator for
the anti-neutrino.]
For the Majorana neutrino either  \(\nu(x)\) or \(\nu^\dagger(x)\) can
annihilate and create a neutrino.
This causes a cancellation.
If one evaluates the matrix element between the  states with
quantum numbers \((\vec p, s)\) and \((\vec p\,', t)\),
it is reduced by
 \((s|\vec p\,| + t |\vec p\,'\,| )/ \mnu\) compared with the one for the
Dirac neutrinos.
This is a manifestation that a Majorana neutrino cannot have diagonal
magnetic moment by the CPT invariance: the moment is necessarily
proportional to the spin  and turns to the opposite direction
under the CPT transformation, while \(\vec B\) stays in the same direction.

\section{Magnetic moments  in a degenerate electron plasma}

The  vertex function \(F^\mu\) in a degenerate
electron plasma is evaluated with \(\dgket\) in Eq.~(\ref{Fmu}),
the state in which all the one
particle states that have momentum below \pF\ are filled.
In the actual calculation, it suffices that we simply use the modified
electron propagators for the Feynman rules \cite{fet}.
There are four nonzero propagators,
which we label with a double index \((ij)\), \(i, j = 1, 2\).
They are defined by
\be
S^{(ij)}(p) \equiv (-i) \int dx e^{ipx} \dgbra \T \psi^{(i)}(x)
\bar\psi^{(j)}(0)
\dgket,
\ee
where \(\psi^{(1)} = \epsilon \ec^\dagger\),  \(\psi^{(2)} =  \e\),
 \(\bar\psi^{(1)} = \e^\dagger\), and   \(\bar\psi^{(2)} = \epsilon
\ec^\dagger\);
the notation takes after the one for a four-component Dirac spinor
(see the Appendix).
They are evaluated  straightforward and obtained as \cite{ada}
\bea
S^{(ij)}(p) = S_{\rm el}^{(ij)}(\vec p\,)
{1\over p^0 - \Ee(\vec p\,) + i\varepsilon }
- S_{\rm ps}^{(ij)}(\vec p\,)
{1\over p^0 + \Ee(\vec p\,) - i\varepsilon } \nonumber\\
- S_{\rm el}^{(ij)}(\vec p\,)
{\theta( \pF - |\vec p\,| ) \over  p^0 - \Ee(\vec p\,) + i\varepsilon }
+ S_{\rm el}^{(ij)}(\vec p\,)
{\theta( \pF - |\vec p\,| ) \over  p^0 - \Ee(\vec p\,) - i\varepsilon }
\label{Sij}
\eea
where
\bea
&&S_{\rm el}^{(11)} = S_{\rm ps}^{(11)} =S_{\rm el }^{(22)} = S_{\rm
ps}^{(22)} = {\me \over 2 \Ee(\vec p\,)} \nonumber\\
&&S_{\rm el}^{(12)} =  - S_{\rm ps}^{(21)} = {\Ee(\vec p\,) + \vec p\cdot\vec
\sigma\over 2 \Ee(\vec p\,)},  \quad
S_{\rm el}^{(21)} =  - S_{\rm ps}^{(12)} = {\Ee(\vec p\,) - \vec p\cdot\vec
\sigma\over 2 \Ee(\vec p\,)}.  \label{Sels}
\eea
We define \(\Fdg^\mu\), the term proper to the degenerate plasma, by
\be \Fdg^\mu (p_1, p_2) \equiv  F^\mu(p_1, p_2) - \Fvc^\mu(p_1, p_2). \ee
Note that the first two terms in Eq.~(\ref{Sij}) is the vacuum contribution
and the last two terms represent the correction due to the plasma.
Thus \(\Fdg^\mu\) is the difference of  two amplitudes  evaluated
with Eq.~(\ref{Sij})  and  with only
the first two terms of the electron propagators.

We first examine the contribution from Fig.~1~(a).
The Feynman integral typically have forms
\be \int^{|\vec p\,| \sim p_{\rm F}} {d^4p\over (2\pi)^4 }
{1 \over [ ( ( p^0 + p_1^0) \pm  \Ee(\vec p + \vec
p_1\,) \pm i\varepsilon]
[ ( ( p^0 - p_2^0) \pm \Ee(\vec p -  \vec p_2\,) \pm i\varepsilon]
[(p^0 )^2 - \EW^2(\vec p\,) + i\varepsilon ] },
\ee
where \(E_W(\vec p\,) \equiv \sqrt{ \vec p\,^2 + m_W^2}\).
Suppose  we do the \(p^0\)-integral first by the contour integral method.
The kinematics for the process we are considering is
\be p_1^0, p_2^0, |\vec p_1\,|, |\vec p_2\,| \ll \me, \pF \ll \mW .\ee
Thus, to leading order in \(1/\mW\), the dominant contributions are
given by picking  the residues at \(p^0 \sim \pm \Ee\).
They are proportional to  \(1/\mW^2\), while the residues at \(p^0\sim
\pm\EW\)
give a contribution proportional to  \(1/\mW^3\) at most.
This means that the W boson propagator \(G_{\mu\nu}(p) \) can be  safely
contracted to the point
form \( g_{\mu\nu}/\mW^2\).  We obtain
\bea
\lefteqn{ \Fdg^\mu(p_1, p_2)  =  \left( {1\over \mW^2 } \right) \int
{d\vec p \over (2\pi)^3}
\sigmabar^\lambda } \nonumber\\
&&\left\{
{\left[  S_{\rm el}^{(21)} (\vec p -\vec p_1\,) \sigmabar^\mu
S_{\rm el}^{(21)}(\vec p + \vec p_2)
+ S_{\rm el}^{(22)} (\vec p -\vec p_1\,) \sigma^\mu
S_{\rm el}^{(11)}(\vec p + \vec p_2) \right]
 \theta(|\vec p-\vec p_1| - \pF) \theta(\pF - |\vec p + \vec p_2|  )\over
q^0 - ( \Ee(\vec p - \vec p_1\,)- \Ee(\vec p + \vec p_2 \,) ) + i\varepsilon}
\right.
\nonumber\\
&&+{ \left[ S_{\rm el}^{(21)} (\vec p -\vec p_1\,) \sigmabar^\mu
S_{\rm ps}^{(21)}(\vec p + \vec p_2)
+ S_{\rm el}^{(22)} (\vec p -\vec p_1\,) \sigma^\mu
S_{\rm ps}^{(11)}(\vec p + \vec p_2) \right]
 \theta( \pF - | \vec p - \vec p_1 | ) \over
q^0 - ( \Ee(\vec p - \vec p_1\,)
+ \Ee(\vec p + \vec p_2 \,) ) + i\varepsilon}
\nonumber\\
&&+ { \left[  S_{\rm el}^{(21)} (\vec p -\vec p_1\,) \sigmabar^\mu
S_{\rm el}^{(21)}(\vec p + \vec p_2)
+ S_{\rm el}^{(22)} (\vec p -\vec p_1\,) \sigma^\mu
S_{\rm el}^{(11)}(\vec p + \vec p_2) \right]
 \theta(|\vec p + \vec p_2 | - \pF) \theta(\pF - |\vec p - \vec p_1|  )\over
- q^0 - ( \Ee(\vec p + \vec p_2 \,)- \Ee(\vec p - \vec p_1 \,) ) +
i\varepsilon}
\nonumber\\
&&+ \left.{\left[ S_{\rm ps}^{(21)} (\vec p -\vec p_1\,) \sigmabar^\mu
S_{\rm el}^{(21)}(\vec p + \vec p_2)
+ S_{\rm ps}^{(22)} (\vec p -\vec p_1\,) \sigma^\mu
S_{\rm el}^{(11)}(\vec p + \vec p_2) \right]
 \theta( \pF - | \vec p + \vec p_2 | ) \over
- q^0 - ( \Ee(\vec p - \vec p_1\,)
+ \Ee(\vec p + \vec p_2 \,) ) + i\varepsilon}
\right\}
\sigmabar_\lambda, \label{Fdg}
\eea
where \(\sigma^\mu = (1, \vec\sigma )\).
By the similar analysis, we realize that the Fig.~1~(b) gives a contribution
proportional to  \(1/\mW^3\) at most and  neglect it.

Each term in Eq.~(\ref{Fdg}) has  definite  physical interpretation.
The first term in the brace represents the  excitation of a electron-hole pair
by photon
and its subsequent annihilation into a neutrino pair by W boson exchange.
The second term denotes  the  vacuum polarization of
a  electron-positron pair where the intermediate electron has a
momentum in the Fermi sphere.
Note  that this process is now forbidden by the Pauli blocking and the
corresponding correction should appear in \(F^\mu_{\rm dg}\).
The third  and forth terms represent   the time-reversed   processes of the
first
and the second, respectively.
Note also \(\Fdg^\mu\) now depends only on \(q = -p_1 - p_2\), as one can
see
easily by a change of variable \(\vec p\, \rightarrow \vec p+\vec p_1\).

We  specify the form of \(\Fdg^\mu\) as we did for \(\Fvc^\mu\) and
identify the form factor for the magnetic dipole moments.
 We have assumed the plasma is isotropic and homogeneous, thus
the temporal component \(F^0_{\rm dg}\) is  a scalar while the spatial
components \(\vec F_{\rm dg}\) constitute a vector under spatial rotations.
There are two structures, \(1\)  and \(\vec\sigma\cdot\vec
q\), for \(F^0_{\rm dg}\) and four structures, \(\vec q\),
\((\vec\sigma\cdot\vec
q\,)\vec q\), \(\vec \sigma\), and \(\vec\sigma\times\vec q\), for
\(\vec F_{\rm dg}\).
An important information is given by
contracting the W propagator  and
applying  ``Fierz'' transformation,
\be (\sigmabar^\lambda)_{\alpha\beta} (\sigmabar_\lambda)_{\gamma\delta}
= - (\sigmabar^\lambda)_{\gamma\beta} (\sigmabar_\lambda)_{\alpha\delta},
\ee
to Eq.~(\ref{Fmu}).
We then have
\bea \Fdg^\mu & = & {i\sigmabar_\nu \over \mW^2}\int dx e^{iqx}
\dgbra\, {\rm T}\,  \e^\dagger(x) \bar\sigma^\nu \e(x) \, \left[ e^\dagger(0)
\bar\sigma^\mu e(0)
- \ec^\dagger(0) \bar\sigma^\mu \ec(0) \right]  \dgket
\nonumber\\
&&-(\mbox{the same term with \(|0\rangle\)}).
\eea
The current \(\e^\dagger(x) \bar\sigma^\nu \e(x) \) is the V-A current of
electron
in the terminology of the Dirac notation.
\(\Fdg^\mu\) is then  divided into two components,
one from  the vector current and the other from the  axial vector
current \cite{ada},
\be \Fdg^\mu = {1\over 2 \mW^2} \left( \Pi^{\mu\nu}_{\rm V}(q)
+ \Pi^{\mu\nu}_{\rm A}(q) \right) \sigmabar_\nu. \ee
The function \(\Pi^{\mu\nu}_{\rm V}\), which have come from the vector
current,
is the polarization tensor of the electromagnetic field in
a medium and consists of two form factors (see ref.~\cite{ada} and references
cited therein).
Only one structure \(\vec\sigma\times \vec q\)  fits to the axial component
\(\Pi^{\mu\nu}_{\rm A}(q)  \sigmabar_\nu\)
as is verified by P-invariance of \(|0\rangle\) and \(\dgket\).
We thus realize that \(F^\mu_{\rm dg}\)  has the form
\bea
F^0_{\rm dg} &=&  f_{\rm l} \left[ |\vec q\,|^2  +q^0  (\vec \sigma \cdot
\vec q\,) \right], \label{Fdg0}\\
\vec F_{\rm dg} & =  & f_{\rm l} \, q^0 \vec q\, \left[  1
+ { q^0 \over |\vec q\,|^2  } (\vec \sigma \cdot
\vec q\,) \right] +  f_{\rm t}\, \left[ |\vec q\,|^2 \vec \sigma
- (\vec\sigma\cdot\vec q\,)\vec q\,)\right] +
f_{\rm m} \, i (\vec \sigma \times \vec q\,)\label{Fdgi}
\eea
with three form factors \(f_{\rm l}\), \(f_{\rm t}\), and \(f_{\rm m}\),
which depend on the rotational invariants \(|\vec q\,|\) and \(q^0\).
The gauge invariance, \(q^0 F^0_{\rm dg} - \vec
q\cdot \vec F_{\rm dg} = 0\), is now easily seen to hold.
\(f_{\rm m}\) at \(q^0 = 0\) is the  magnetic dipole moment,
\be\mu_{ab} = { e g^2 \over 2} V_{\nu_a {\rm e}} V^*_{\nu_b {\rm e}}
f_{\rm m}. \ee
\(f_{\rm l}\), \(f_{\rm t}\), and \(f_{\rm m}\) are
proportional to \({\cal T}_L\), \({\cal T}_T\), and \({\cal T}_P\)
in Ref.~\cite{nie}, respectively.

Since the state \dgket\ is T-invariant, the same equation as (\ref{Tinv})
applies to
\(\Fdg^\mu(q)\) with \newline \(q =- p_1-p_2\).
This  shows that all the form factors in Eqs.~(\ref{Fdg0}) and (\ref{Fdgi})
are even functions of \(q^0\).
This result especially indicates that the electric dipole moment,
the coefficient of \((\vec\sigma\cdot\vec q\,)\) in \(\Fdg^0\) at \(q^0= 0\),
is zero up to  one loop correction even in degenerate electron gas.

CP transforms \dgket\ to the state  of degenerate positrons.
It leads us to a relation
\be F^\mu_{\rm dg, ps}(q) = -\epsilon \bar \Fdg^\mu{}^{\rm T}(\bar q)
\epsilon
\label{Fdgps},\ee
where \(F_{\rm dg,ps}^\mu\) is the counterpart of \(\Fdg^\mu\)
evaluated for the degenerate positrons.
The structure \((\vec\sigma\times\vec q\,) \) is odd under the operation
defined by the right hand side of (\ref{Fdgps}), while the other structures
are even.  Thus the positron-induced magnetic moment has an  opposite sign
to electron-induced one.
This also indicates that electron and positron contribute subtractively
if they coexist in a plasma.
We also  infer that the induced magnetic moment vanishes in
a plasma where the temperature is much higher than the chemical potential
of electrons.
These inferences are consistent to the explicit formula obtained in
Refs.~\cite{nie} for  \({\cal T}_P\).

In order to obtain \(f_{\rm m}\),
we evaluate the trace \(I \equiv {\rm Tr}(\sigma^j F^i_{\rm dg})/2 \)
and retain only the terms that are proportional to the anti-symmetric tensor
\(\epsilon_{ijk}\).
Using Eqs.~(\ref{Sels}), (\ref{Fdg}), and (\ref{S3}) in the Appendix, we found
\bea I& =&  i \epsilon_{ijk} \left( {1\over  \mW^2}\right) \int {d\vec p
\over
(2\pi)^3}\nonumber\\
&& \times\left[
\left( { p^k + q^k \over \Ee(\vec p + \vec q\,)}- { p^k \over \Ee(\vec p\,)}
\right)
{\theta( | \vec p + \vec q\,| - \pF ) \theta(\pF - | \vec p\,| ) (\Ee(\vec p +
\vec
q\,)- \Ee(\vec p\,))
\over (q^0)^2- (\Ee(\vec p + \vec q\,)- \Ee(\vec p\,))^2 }\right.\nonumber \\
&&- \left. \left( { p^k + q^k \over \Ee(\vec p + \vec q\,)}
+ { p^k \over \Ee(\vec p\,)} \right) {\theta(\pF - |\vec p + \vec q\,|)
(\Ee(\vec p + \vec q\,)+ \Ee(\vec p\,))
\over (q^0)^2- (\Ee(\vec p + \vec q\,) + \Ee(\vec p\,))^2 }\right].
\label{I}\eea
Note that no absorbtive part is induced for \(q^0 = 0\).
The first integrand in the bracket in Eq.~(\ref{I})  is  obviously nonzero
only in the shell-like region at the surface of the Fermi sphere with the
width
\(|\vec q\,|\) (see Fig.~3).
The integral of the second term in the bracket can also be evaluated
in the same region:
The second integrand, except for the factor \(\theta(\pF - |\vec p + \vec
q\,|)\),
is odd under \(\vec p\rightarrow -\vec p - \vec q\); thus we insert
\(1 = \theta(\pF - |\vec p \,|) + \theta( |\vec p \,| - \pF )\) into the
integrand,
neglect the part from \(\theta(\pF - |\vec p \,|)\), and change the integral
variable as \(\vec p\rightarrow -\vec p - \vec q\).
We approximate the integral measure to
\be d\vec p = 2\pi \pF^2 |\vec q\,| \sin \phi \cos\phi \, d\phi, \quad
0\leq\phi\leq
{\pi\over 2}, \ee
where \(\phi\) is the angle between \(\hat q\) and \(\hat p\).
The integral \(I\) is carried out easily to leading order in \(|\vec
q\,|/\EF\),
where \(\EF\) is the Fermi energy of the electrons.
We found \( I \simeq (-i) \epsilon_{ijk} q^k ( \pF/ 4\pi^2 \mW^2)\), or
\be f_{\rm m } = - {1\over 4\pi^2} {\pF\over \mW^2}. \label{fm}\ee
The magnetic dipole moments are given by
\be \mu_{ab} = - {e \GF \pF \over \sqrt 2 \pi^2 }V_{\nu_a \rm e}
V^*_{\nu_b \rm e} \label{mudgab}\ee
or numerically
\bea \mu_{ab} &= & -8.6 \times 10^{-13} \muB \left(V_{\nu_a \rm e}
V^*_{\nu_b \rm e} \right) \left( {\pF\over 1\mbox{ MeV}}\right)
\nonumber\\
& = &
-4.4 \times 10^{-15} \muB \left(V_{\nu_a \rm e} V^*_{\nu_b \rm e} \right)
\left( Y_{\rm e}\,\rho( {\rm g/cm^3})\right)^{1/3}.  \eea
In fact the neutral current interaction, which we have been neglecting,
gives a contribution to the diagonal component.
Its magnitude relative to the charged current contribution is
readily obtained by comparing the coupling constant of its effective
four Fermi interaction with the one from the charged current.
The correction is made by replacing
 \(|V_{\nu_a \rm e}|^2 \) with \((|V_{\nu_a \rm e}|^2 - 1/2)\) in
Eq.~(\ref{mudgab}).
Note that the GIM cancellation does not work for the transition moments
since only electron can contribute to \(F^\mu_{\rm dg}\).

We compare the one-loop induced vertices in the vacuum,
Eqs.~(\ref{Fvc}) and (\ref{fM}), and in the electron background,
Eqs.~(\ref{Fdgi}) and (\ref{fm}), and notice that
there are two places where the magnetic moments are significantly
enhanced by degenerate electrons.
One place is where the neutrino has non-relativistic momenta;
whether we consider Dirac or Majorana neutrino,  helicity flipping
or non-flipping process,
the dipole moment is enhanced by the factor (8\pF/3\mnu)
when compared with the vacuum case.

The other place is for flavor-changing processes.
The induced vertex has the form Eq.~(\ref{Lm}) in the standard model
and the helicity-flipping process for a relativistic neutrino
is suppressed by the  factor \(\mnu/E_\nu\) as one can see using
the expansion (\ref{nuex}).
There is no such suppression for the helicity-conserving one.
The  moments \(\mu_{ab}\) can induce a transition
between \(\nu_{\rm e}\) and \(\nu_\mu\) (or \(\nu_\tau\)) in the presence
of magnetic field with their helicity intact.
We should mention one caution here.
When the neutrino propagates in a dense medium,
its energy is changed by the coherent interactions with the medium
\cite{wol}.
If this effect is stronger than the one from the mass matrix,
the energy eigenstate becomes the interaction eigenstate
instead of the mass eigenstate.
For the interaction eigenstate,  the one-loop
induced magnetic moments can no longer generate
``flavor-changing'' processes.
The condition that this will not happen is that the mass difference is
big enough, i.e.  \(\Delta m^2/E_\nu  \gg 2\sqrt 2 \GF n_{\rm e}\),
or numerically%
\footnote{This condition gives, for example, \(\Delta m \gg 100 \) eV
for \(\rho \sim 10^{10} {\rm g/cm^3}\) and \( E_\nu \sim \pF \sim 10\)
MeV.}
\be {\Delta m^2 \over ( 1 {\rm eV})^2} \gg 1.5 \times 10^{-7} \left(E_\nu\over
1
\rm
MeV\right) Y_{\rm e}\,\rho( {\rm g/cm^3}), \ee
where \(\Delta m^2\) is the squared-mass difference of the neutrinos in
interest.
Provided that this condition is satisfied,
the flavor-changing process is enhanced relative to the vacuum case
because of the absence of
the GIM suppression factor \(\me^2/\mW^2\).

The enhancement is not large enough to affect the solar neutrino flux.
The required magnitude is estimated to be \( \sim 10^{-10} \muB \)
\cite{vol,lim}, while
 \pF\  has the magnitude of \(10^{-2}\) MeV
even at the center of the Sun and the resulting dipole moment is too small.

The  flavor-changing transition
can affect the neutrino flux from a supernova \cite{kik}.
External magnetic field can give a nonzero
momentum transfer to a neutrino traveling in it
if its configuration has  spatial variation.
This gives a  possibility for the flavor-changing
transition which is  different from the one such as the MSW effect \cite{wol}
where the  electron density is not thought to affect
the neutrino  momentum strongly.
In order to establish  a realistic picture for the  flavor-changing process
induced by the magnetic dipole moments,
it is necessary to know the profiles of the magnetic field
and the electron density in the stellar interior
and it is a subject of further study.

\begin{center}
\bf Acknowledgements
\end{center}
The author thanks A.~Riotto  for the discussions and
for letting him know the related papers on the subject.
He also thanks C.~S.~Lim for helpful discussions.

\appendix
\section{Appendix}

In this appendix we related the formula in the two component notation
to those familiar in the four-component Dirac notation.
A  Dirac spinor  and the gamma matrices have two-component decomposition
as
\be \psi = \left(\begin{array}{c}
 \epsilon \ec^\dagger \\ \e \end{array} \right)
\quad  \bar\psi \equiv \psi^\dagger \gamma^0 = ( \e^\dagger,
\epsilon \ec ), \ee
and
\be
\gamma^\mu = \left(\begin{array}{cc} 0 &  \sigmabar^\mu \\ \sigma^\mu &
0
\end{array}\right).\ee
With these definitions,
the Lagrangian (\ref{QED}) is written in the familiar form
\be
{\cal L}_{\rm e} =
\bar\psi \{ i (\partial_\mu - ie A_\mu ) \gamma^\mu - m \} \psi. \ee
The following matrix relation is useful to prove the invariance under
the discrete symmetries, C, P, and T.
\bea
\epsilon \sigma^\mu{}^{\rm T} \epsilon = - \sigmabar^\mu  & \quad &
\epsilon \sigmabar^\mu{}^{\rm T} \epsilon = - \sigma^\mu \nonumber\\
\epsilon \sigma^\mu{}^* \epsilon = - \sigmabar^\mu  & \quad &
\epsilon \sigmabar^\mu{}^* \epsilon = - \sigma^\mu.
\eea
The products of \(\sigma^\mu\) reduce to  expansions,
\bea
\sigmabar^\mu \sigma^\nu \sigmabar^\lambda  & = &
g^{\nu\lambda}\sigmabar^\mu  + g^{\mu\nu}\sigmabar^\lambda -
g^{\mu\lambda}\sigmabar^\nu - i \epsilon^{\kappa\mu\nu\lambda}
\sigmabar_\kappa
\nonumber\\
\sigma^\mu \sigmabar^\nu \sigma^\lambda  & = &
g^{\nu\lambda}\sigma^\mu  + g^{\mu\nu}\sigma^\lambda -
g^{\mu\lambda}\sigma^\nu + i \epsilon^{\kappa\mu\nu\lambda}
\sigmabar_\kappa. \label{S3}
\eea

\newpage
\begin{center}
\bf Figure captions
\end{center}
\begin{description}
\item[Fig. 1:] The Feynman diagrams for the magnetic dipole moments of a
neutrino induced by the charged current interaction.
\item[Fig. 2:] The Feynman diagram with the neutral current interaction
that also contributes to the diagonal component of  the magnetic moments
in a degenerate electron plasma.
\item[Fig. 3:] The cross section of the Fermi sphere. The shadowed region is
the element of the integral region that contributes to the integral \(I\),
Eq.~(\ref{I}).
\end{description}

\end{document}